\documentclass[aps,prb,twocolumn,showpacs,superscriptaddress]{revtex4}
\usepackage{mathrsfs}
\usepackage{verbatim}
\usepackage{epsfig}
\usepackage{subfigure}
\usepackage{graphicx}
\usepackage{amsmath}
\usepackage{amsfonts}
\usepackage{amssymb}
\usepackage{dcolumn}
\usepackage{bm}

\begin{document}
\title{The role of geometry and topological defects in the 1D zero-line modes of graphene}
\author{Xintao Bi}
\affiliation{ICQD, Hefei National Laboratory for Physical Sciences at Microscale, and Synergetic Innovation Centre of Quantum Information and Quantum Physics, University of Science and Technology of China, Hefei, Anhui 230026, China}
\affiliation{CAS Key Laboratory of Strongly-Coupled Quantum Matter Physics and Department of Physics, University of Science and Technology of China, Hefei, Anhui 230026, China.}
\author{Jeil Jung}
\affiliation{Department of Physics, University of Seoul, Seoul 130-742, Korea}
\author{Zhenhua Qiao}
\email[Correspondence author:~~]{qiao@ustc.edu.cn}
\affiliation{ICQD, Hefei National Laboratory for Physical Sciences at Microscale, and Synergetic Innovation Centre of Quantum Information and Quantum Physics, University of Science and Technology of China, Hefei, Anhui 230026, China}
\affiliation{CAS Key Laboratory of Strongly-Coupled Quantum Matter Physics and Department of Physics, University of Science and Technology of China, Hefei, Anhui 230026, China.}
\date{\today}

\begin{abstract}
  Breaking inversion symmetry in chiral graphene systems, \textit{e.g.},
  by applying a perpendicular electric field in chirally-stacked rhombohedral multilayer graphene or
  by introducing staggered sublattice potentials in monolayer graphene, opens up a bulk
  band gap that harbors a quantum valley-Hall state.
  When the gap size is allowed to vary and changes sign in space, a topologically-confined
  one-dimensional (1D)
  zero-line mode (ZLM) is formed along the zero lines of the local gap.
  Here we show that gapless ZLM with distinguishable valley degrees of freedom K and K$'$
  exist for every propagation angle except for the armchair direction that exactly superpose the valleys.
  We further analyze the role of different geometries of top-bottom gated device setups that can be realized in
  experiments, discuss the effects of their edge misalignment,
  and analyze three common forms of topological defects that could
  influence the 1D ZLM transport properties in actual devices.
\end{abstract}
\maketitle

\section{Introduction}
Graphene, a two-dimensional honeycomb-lattice structure of carbon, is intrinsically a zero-gap semimetal\cite{castroneto}.
Its internal binary degrees of freedom such as the valleys, sublattices,
and top/bottom layers in multilayers leads to a two dimensional electron gas
with valley selective chirality of the electrons near the Fermi energy.
As a consequence, breaking the spatial inversion
symmetry through staggered AB sublattice potentials in graphene,
or by applying a vertical electric field in rhombohedral multilayers ~\cite{semenoff,hongki}
opens up a band gap that embodies a bulk valley-Hall effect \cite{QVHE1-DiXiao}
which is robust against disorder provided that the  K and K$'$ valley coupling is weak \cite{robusthall}.
A variety of topologically distinct ground-states result in the presence of
valley selective mass terms in the Dirac Hamiltonian of graphene
\cite{QSHE1,QSHE2,QAHE1,QAHE2,QAHE3,QAHE4,QVHE2-DunghaiLi,QVHE3-Jung,
QVHE3-ZhangFan,QVHE4-Xiao,QVHE5-Qiao,robusthall,QVHE1-DiXiao}
and Dirac fermion zero energy modes \cite{jackiwrebbi,volovik}
are expected when the local gap or mass of the Dirac Hamiltonian changes sign in real space.
These zero modes have been proposed in solitons at domain walls of alternating
single-double bonds of the carbon atoms in polyacetylene~\cite{polyacetylene}
whereas one dimensional (1D) zero-lines of the local Dirac gap in real space
was proposed in bilayer graphene at regions where the applied perpendicular electric field changes sign~\cite{morpurgo,kinkstate-jeil,highway,fpeeters}.
The 1D zero-line modes (ZLM)\cite{partition} due to kinks in the Dirac mass
are naturally expected in other graphenic honeycomb systems like monolayer graphene~\cite{Semenoff,YaoWang},
graphene superlattices~\cite{killi_superlattices}, multilayer graphene~\cite{kinkstate-jeil} and AB/BA stacking domain
at opposite mass honeycomb lattice interfaces \cite{nanoroad},
or tilted multilayer grain boundaries\cite{eunah,pnasfan}.
It is expected from theory that these ZLMs will have special transport properties compared to
ordinary 1D channels.
For instance, the analysis of the low energy Hamiltonian indicates that the ZLMs associated
with different valleys are time-reversal opposites encoded with
opposite chirality that propagate along the same trajectory, and therefore have a substantial
spatial overlap that would enhance their mixing in the presence of disorder.
However, analyses based on a realistic tight-binding Hamiltonian have shown that the ZLMs
are extremely robust against backscattering for both short and long-range disorder,
exhibiting long mean-free paths on the order of hundreds of micrometers for typical magnitudes
of disorder strength expected in experiments.~\cite{highway,nanoroad}
The reason for this unexpected robustness to backscattering has been attributed to the rather wide spread of the
wave-functions across the zero-line that suppresses the backscattering probability, resulting in
practically dissipationless transport along any curved zero-line trajectory.
Interesting transport features are expected in more complex geometries formed by intersecting zero-lines~\cite{partition}
in beam splitter geometries that dictate special current partition laws in transport channels in a network of ZLMs.
These intersecting zero-line setups can in principle be realized by means of
spatially-tunable electric fields in bilayer graphene or by a uniform field
in systems with multiple stacking-fault domains~\cite{mceuen}.
Despite that the ZLMs in graphene have been theoretically proposed for several years
the experiments have shown a rather limited progress in their realization until a recent observation and
transport measurement of a zero-line in a bilayer graphene deposited on SiO$_2$ displaying a stacking-fault domain wall~\cite{NatureZLM}.
It is desirable that the ZLM can be designed in a more systematic manner in experiments,
and the success in realizing them in gate tunable devices depends primarily on the degree of precision achievable
in the alignment of the top-bottom electric gates used to generate the local band gap of bilayer graphene,
as well as the achievable quality of the crystal.

In this article we address the question on the feasibility of experimental realization of ZLM in gate tunable devices
by providing a detailed account of the adverse effects that can be introduced in
the ZLMs by the misalignment of gates and topological lattice defects.
To this end we calculate the electronic structure and the two-terminal electron transport for a number of device
geometries and short-range disorder configurations for bilayer and
single layer ribbon geometries to exemplify general behaviors of the ZLMs.
The article is organized as follows.
In Sec.~\ref{HamiltonianandMethods} we introduce the system Hamiltonian and the transport calculation methods.
Then in Sec.~\ref{SectionIII} we move on to explore the effects of device geometry in the ZLMs and propose
a new simplified scheme of global top-bottom gates intercalated with smaller gates
that can improve the chances of detecting the ZLMs in experiments.
In Sec.~\ref{SectionIV}, we examine the robustness of the ZLMs in the presence of three different kinds of topological defects.
Finally in Sec.~\ref{SectionV} we close the paper with the summary and conclusions.

\section{Hamiltonian and methods}\label{HamiltonianandMethods}
The  ZLM solitons can appear in AB-stacked bilayer graphene in the
presence of spatially-varying interlayer potential difference.
The corresponding $\pi$-orbital tight-binding
Hamiltonian in real-space representation can be written as:
\begin{eqnarray}
H =   \sum_{\langle i,j \rangle}  t_{i,j} \,\,  c_i^{\dag} c_j
+ \sum_{i} U_{i} \, \, c_{i}^{\dag} c_{i}, \label{eq1}
\end{eqnarray}
where the first term denotes both the intralayer and interlayer nearest-neighbor hopping of the electrons,
with $t_{i,j}$ being the magnitude of in-plane hopping with a value of $-2.6~{\rm eV}$ or a
vertical inter-layer hopping amplitude with a value of $0.36~{\rm eV}$ \cite{tbwannier}.
Here $c^{\dag}_i$ and $c_{i}$ are respectively the creation and annihilation operators at site $i$.
The second term measures the interlayer potential difference, with $U_{i}$ being the site potential energy.

The transport properties of the ZLMs are numerically simulated by employing a mesoscopic two-terminal device,
and the conductance $G_{\rm RL}$ from the left to right terminal can be evaluated using the Landauer-B\"{u}ttiker formula:~\cite{datta}
\begin{equation}
G_{\rm RL}=\frac{e^2}{h}~{\rm Tr} (\Gamma_{\rm R} {\rm G}^r \Gamma_{\rm L} {\rm G}^a).
\end{equation}
where ${\rm G}^{r,a}$ are the retarded and advanced Green functions of the central scattering region.
The quantity $\Gamma_{\rm L/R}$ is a linewidth function describing the coupling between
$L/R$-terminal and the scattering region and can be obtained through the self-energy $\Sigma^{r}_{\rm L/R}$ using a transfer-matrix method~\cite{transfer}.
Finally, the propagation of the ZLM from the $L/R$-terminal is visualised by presenting the the local density of states (LDOS) at each site:\cite{datta}
\begin{equation}
\rho_{\rm L/R}({\bf r}, \varepsilon)=\frac{1}{2\pi}[{\rm G}^r \Gamma _{\rm L/R}
{\rm G}^a]_{{\bf r} {\bf r}},
\end{equation}
where $\varepsilon$ is the Fermi energy.

\begin{figure}
\includegraphics[width=8.5cm,angle=0]{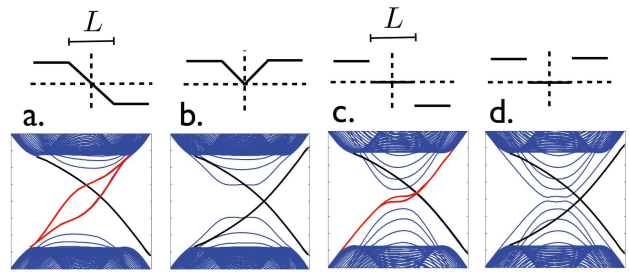}
\caption{
Bands corresponding to zero-line (red), valley Hall (black), and
the interface bound modes (blue) for distinct profiles of the interlayer potential difference.
The presence of interlayer potential variations indicated in the upper panel causes the band gap closure
due to the formation of interface bound modes (IBM). The comparison between a.-b. and c.-b. panels shows
the specific dependence of the IBM on different shapes of the mass variation at the interface,
while that between a.-c. and b.-d. panels indicates the essentialness of the mass reversal
for the formation of the ZLM.
Figure adapted from the supplementary information of Ref.~[\onlinecite{highway}].
}\label{figibm}
\end{figure}

\section{Device geometry effects in the zero-line modes} \label{SectionIII}
The ZLMs in bilayer graphene have been theoretically proposed to appear along the zero lines of the interlayer potential difference.
In the absence of an AB/BA stacking fault that reverse the valley-Hall conductivity \cite{eunah,nanoroad,pnasfan},
it is necessary to reverse the sign of the electric fields in order to produce such a zero field line.
Usually, the confinement effect caused by the interface potentials will partially localise the wave functions
of the bulk modes and lead to the formation of the interface bound modes (IBM)~\cite{morpurgo,highway,fpeeters}, whose physical origin
is qualitatively different from that of the ZLM. The energy bands of the IBM depend on the details of the potential profiles and will
reduce the band gap where the ZLM is hosted (see Fig.~\ref{figibm}). To minimize the band-gap closure, a sharp potential variation is desired.
%while keeping the the sharpest potential variation
%is desired in order to minimize the band-gap closure due to the formation of partially localized interface bound modes (IBM)
%that evolve from the bulk modes by partial wave function localization
%due to confiment effects by the interface potentials~\cite{morpurgo,highway,fpeeters}.
%The energy bands of the IBM depend on the
%details of the potentials and progressively reduce the band gap
%and has a physical origin that is qualitatively different from the ZLM (see Fig.~\ref{figibm}).

\begin{figure}
  \centering
  \includegraphics[width=6.5cm,angle=0]{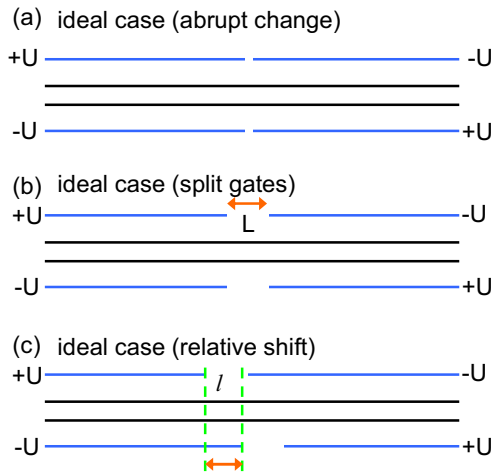}
  \caption{
  Model system setups of a bilayer graphene in the presence of spatially varying interlayer potential difference.
  In actual devices the potential profile variation depends on the thickness of the dielectric spacer and the in-plane alignment
  between the gate edges. (a) The interlayer potential difference changes sign abruptly for perfect top-bottom gate alignment.
  (b) A finite in-plane separation $L$ between two top (bottom) gates smoothens the potential variation.
  (c) A relative shift $l$ which introduces misalignment between top and bottom gates further reduces
  the abruptness of the potential variation.}
  \label{FIG3}
\end{figure}

For this purpose two pairs of top and bottom gates adjacent to each other are needed, as shown in Fig.~\ref{FIG3}.
Ideally, the two sets of top and bottom gates should be arranged in a perfect way as displayed in Fig.~\ref{FIG3}~(a) to induce an abrupt change of the electric field.
Nevertheless, the error bars of current lithographic techniques is expected to introduce small misalignments between the gate edges
on the order of a few tens of nanometers.
In what follows in this section we analyze the effects of device geometry such as the zero-line orientation angle and the misalignment of the gate edges.
We begin by showing in subsection~\ref{SectionIIIA} that the ZLMs have identifiable valley indices for the counter-propagating edge modes along arbitrary
zero lines except the armchair direction.
In subsection~\ref{SectionIIIB}, we focus on two typical misalignments between the adjacent gates in the domain wall and discuss the
consequence of their existence on the dispersion relation of ZLMs.
Finally in subsection~\ref{SectionIIIC}, we propose an alternative device setup which reduces the complexity
in the gate alignment process and simplifies the need for a precise control in patterning the electric gates
required in the experimental realization of the ZLMs.

\begin{figure}
\includegraphics[width=8.5cm,angle=0]{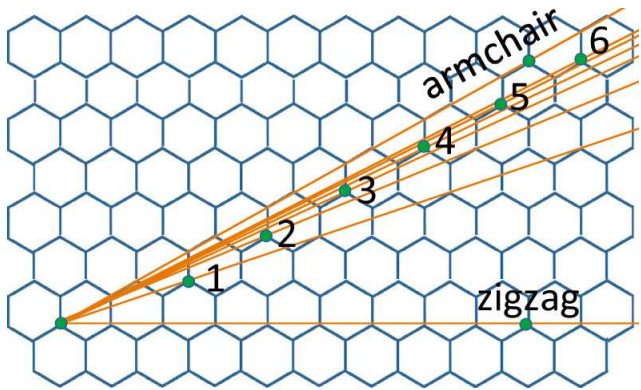}
\caption{
Six intermediate orientation angles between the horizontal zigzag edge
and the armchair edge angle of $\pi/6$.
Nanoribbon geometries with edges oriented along the intermediate angles
are labeled with indices $1-6$ and have ribbon supercells of different periodicity.
}
\label{6orientations}
\end{figure}

\subsection{Zero-line modes with arbitrary orientations}\label{SectionIIIA}
Up to the present, two representative graphene ribbons have been broadly investigated in the study of the ZLMs: the zigzag ribbon where valleys are well-separated and
the armchair ribbon where the K and K$^\prime$ valleys exactly overlap and are therefore indistinguishable.
Here we focus on the electronic structure of the valley Hall modes and the kink ZLMs when the orientation of the ribbons deviate from both zigzag and armchair directions.
\begin{figure}
\includegraphics[width=8.5cm,height=20.5cm,angle=0]{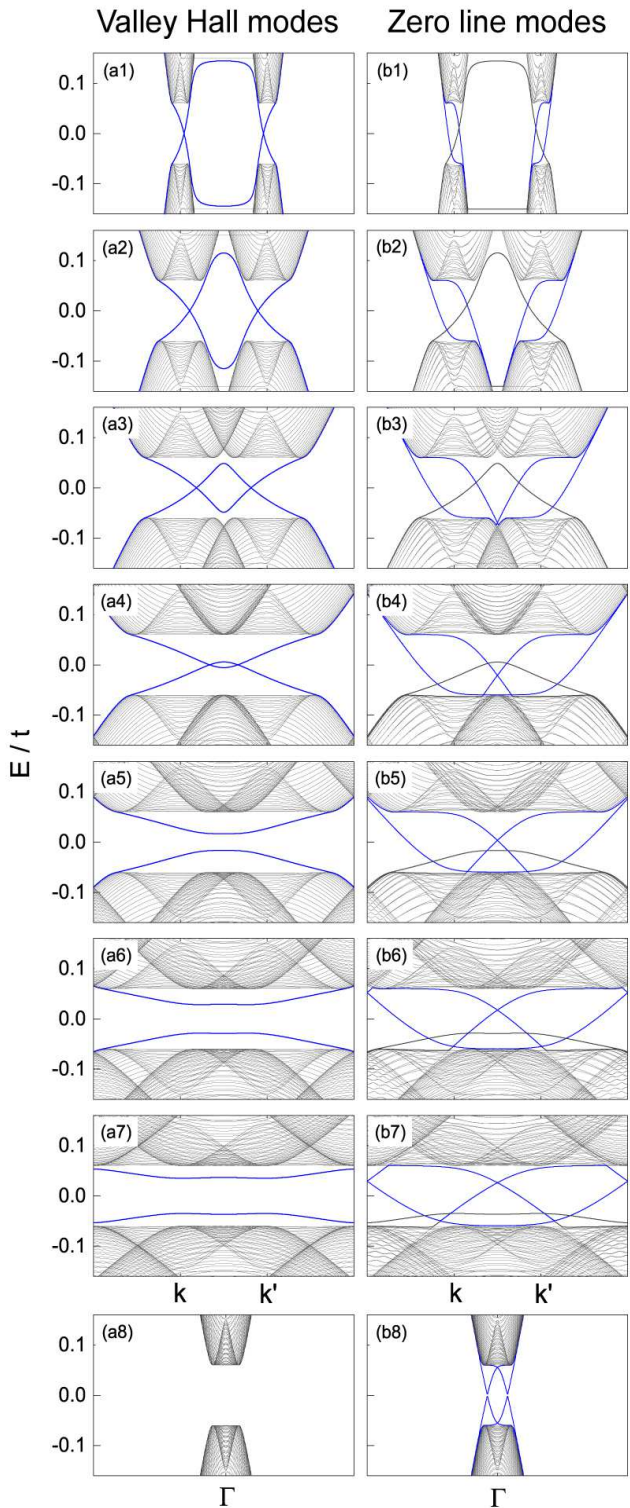}
\caption{Band structure of quantum valley Hall modes and the corresponding ZLMs for different bilayer graphene ribbons given in Fig.~\ref{6orientations}. The QVH edge modes and the ZLMs are labelled in blue. K and K$^\prime$ denote the two valleys which are totally mixed at the $\Gamma$ point for the armchair ribbon. (a1) - (a8) Bandstructure of quantum valley Hall modes. (b1) - (b8) Bandstructure of ZLMs.}
\label{band-6orientations}
\end{figure}
We select eight representative orientations with different ribbon periodicities including the zigzag,
armchair and six other orientations labeled with (``1"-``6") that are neither zigzag nor armchair,
as shown in Fig.~\ref{6orientations}.
The associated band structures of the valley Hall modes are shown in
Figs.~\ref{band-6orientations}(a1)-\ref{band-6orientations}(a8) with the interlayer potential difference of $2U=1.0t$.
It can be seen that the bulk band gaps $\Delta$ are identical for all kinds of bilayer graphene ribbons,
and various gapless and gapped edge modes depicted in blue lines appear for each orientation except the armchair case.
In Fig.~\ref{band-6orientations}~(a1) for the zigzag bilayer graphene ribbon, two pairs of gapless edge modes located at valleys K and K$^\prime$ are well separated in momentum space.
The valley-Hall edge modes corresponding to valleys K and K$^\prime$ are connected with the same valence (conduction) bands,
indicating they are intrinsically distinct from the gapless edge modes in the quantum Hall insulators and $\mathbb{Z}_2$ topological insulators.
When the orientation gradually deviates from the zigzag direction, the two valleys approach each other while the
edge modes remain gapless in Figs.~\ref{band-6orientations}~(a2)-\ref{band-6orientations}~(a4) for the ribbons labelled as ``1"-``3" in Fig.~\ref{6orientations}.
When the ribbons further evolve towards the armchair profile (see the zero lines labeled ``4"-``6" in Fig.~\ref{6orientations}),
the valley-Hall edge modes in Figs~\ref{band-6orientations}~(a5)-\ref{band-6orientations}~(a7) gradually develop gaps.
This behavior can be understood as resulting from the inter-valley scattering due to the mixing of the two valleys.
In the limit of maximum inter-valley scattering for the armchair ribbon the edge modes vanish completely, as shown in Fig.~\ref{band-6orientations}~(a8).
The band evolution of the QVH states for our given ribbon sequence implies that the vanishing of the valley-Hall modes
undergoes a gradual process accompanied with continuous variation of the edge state gaps,
which in turn reflects the gradual increase in the strength of inter-valley scattering.

A contrasting behavior is delivered on the right column of Fig.~\ref{band-6orientations}, which
plot the band structures of the ZLMs under spatially varying interlayer potential differences with $|U|=0.5~t$, as
illustrated in Fig.~\ref{FIG3}~(a). The bulk band gaps $\Delta$ remains the same as in the case with uniform interlayer potential difference.
Inside the bulk band gaps, the ZLMs are highlighted in blue and appear for all the eight kinds of ribbons. For the zigzag zero line displayed in Fig.~\ref{band-6orientations}~(b1), two pairs of gapless ZLM with opposite propagation directions are well separated and located at different valleys in the momentum space. When the zero line deviates from the zigzag direction, as plotted in Figs.~\ref{band-6orientations}~(b2)-\ref{band-6orientations}~(b7), the two pairs of gapless ZLMs gradually cross each other, accompanied by the neighboring valleys K and K$^\prime$.
When the zero line evolves into the armchair direction, the two pairs of ZLMs completely overlap in momentum space and a small avoided crossing gap opens up (See Fig.~\ref{band-6orientations}(b8)).
A striking difference from their QVH edge mode counterparts is that the ZLMs remain all the way
gapless by connecting the conduction and valence bands, except for the aforementioned tiny avoided gap for the armchair case.

It is noteworthy that as the bilayer graphene ribbon gradually deviates from the zigzag direction
to the series of ribbon geometries listed in Fig.~\ref{6orientations} the valleys K and K$^\prime$ remain distinguishable
while their separation in momentum space becomes smaller until they merge completely for the armchair ribbon configuration
Therefore, the arbitrariness in the propagation direction of the zero lines does not pose a severe challenge for the experimental realization of ZLMs.
Based on earlier transport simulations~\cite{highway}, we expect that the ZLMs should be robust against moderate disorder, both long- and short-range,
due to the wide spread of the ZLM wavefunctions.

\subsection{Misalignment of gates}\label{SectionIIIB}

Even though many different types of theoretical studies of the ZLM have been presented in the literature
there has been only a limited progress towards their experimental realization \cite{NatureZLM}.
Most theoretical investigations of the ZLMs have been based on the ideal limit where the potential profile
between the layers changes abruptly across the zero line as shown in Fig.~\ref{FIG3}(a).
In actual preparations, however, deviations due to the finite separation between the gate and graphene,
and the in-plane misalignment of the gates are unavoidable.
Two sources of misalignment are illustrated schematically in Figs.~\ref{FIG3}(b) and \ref{FIG3}(c)
that obscures the formation of ZLMs.
Below we discuss these negative factors in detail, first assuming an extreme case where the interlayer potential difference is zero
all along across the lateral extent of the zero lines
and later switching to a more realistic profile where the interlayer potential difference varies linearly~\cite{highway}.
\begin{figure}
  \includegraphics[width=8.5cm,angle=0]{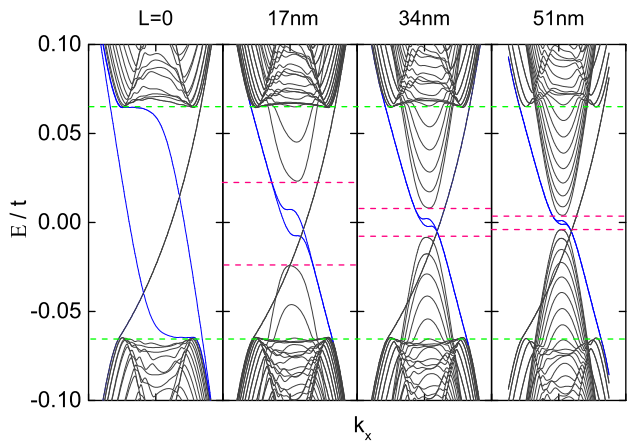}
  \caption{Band structure (only shown the bands around valley K) of a bilayer graphene in the presence of spatially varying interlayer potential difference for gate splitting setups shown in Fig.~\ref{FIG3}(b). L=0, 17, 34, 51~nm respectively in the four panels. The ZLMs are depicted in blue. The green dash lines indicate the bulk band gap hosting the ZLM in the ideal setup of L=0, and the pink dash lines indicate the effective band gaps hosting the ZLMs in setups with increased L.}\label{bands-comparison-constant}
\end{figure}

\begin{figure}
\includegraphics[width=8.5cm,angle=0]{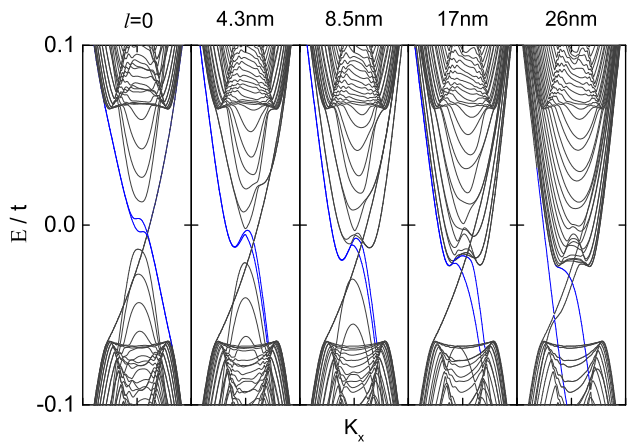}
\caption{Band structure of a bilayer graphene near valley K in the presence of spatially
varying interlayer potential difference for setups shown in Fig.~\ref{FIG3}(c). The ZLMs are depicted in blue.
The separation distance are set to be L=26~nm with} relative shifts of $l=0, 4.3, 8.5, 17, 26$~nm respectively in the five panels.\label{bands-relativeShift}
\end{figure}

Fig.~\ref{FIG3}(b) shows a typical device setup where two precisely aligned top (bottom) gates are separated by a distance of $L$.
Fig.~\ref{bands-comparison-constant} plots the associated band structure near valley K in the presence of an extreme interlayer
potential variation as mentioned before. The separation distances are set to be $L=0,~17,~34,~51$~nm respectively in the four panels of Fig.~\ref{bands-comparison-constant} and the ZLMs are depicted in blue.
It is clear that when $L$ increases additional IBM states appear within the bulk band gap
(indicated by the green dash lines)
which leads to smaller effective band gaps (indicated by the pink dash lines) that encloses the ZLMs.
The finite splitting of gates thus are unfavored in the realization of the ZLMs.
Nevertheless, our numerical results show that even for $L>100$~nm,
sizeable band gaps persist~\cite{highway} and it should be possible to measure the ZLMs in experiments.

Another geometry distortion consists in a vertical misalignment between the top gates and the bottom gates, which is measured in
the relative shift of $l$ as shown in Figs.~\ref{FIG3}(c).
Fig.~\ref{bands-relativeShift} demonstrate the band evolution for a fixed in-plane gate separation distance of $L=26$~nm and different shifts of $l=0$, 4.3, 8.5, 17, and 26~nm.
As $l$ increases, the bound states emerging from the
conduction and that emerging from the valence bands become more and more unbalanced, \textit{i.e.},
the density of the bound states connected with the conduction bands become higher,
while those connected with the valence bands become smaller or even vanishing in the limit of $l \rightarrow L$, although the gapless ZLMs are still located within the bulk band gap. The presence of vertical misalignment between the top and the bottom gates in real devices can become a bottleneck in the experimental realisation of the ZLMs.

\subsection{ Simplified dual gate device setup }\label{SectionIIIC}
As seen from the above discussion the realization of ZLMs is constrained
by both the finite splitting between the two top (bottom) gates
and the vertical misalignment between the top gates and the bottom gates.
Both geometry distortions are difficult to eliminate with present device fabrication capabilities
where the precise alignment of four independent gates (two top and two bottom gates) on the nanometer scale is required.
In this subsection we propose an alternative device setup which can simplify the would-be formidable gate alignment procedure.

\begin{figure}
\includegraphics[width=8.5cm,angle=0]{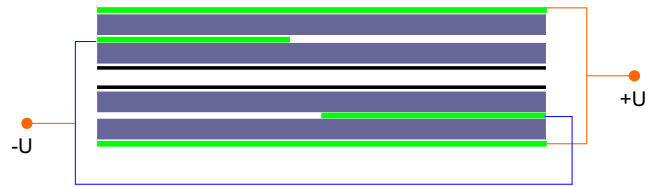}
\caption{
Proposed device setup to measure the 1D ZLM in bilayer graphene. The black solid lines represent the bilayer graphene and
green lines represent the gate electrodes. The dielectric barrier materials are depicted in dark blue.
In this setup, only the two local gate electrodes need to be aligned.} \label{ProposedSetup}
\end{figure}

As shown in Fig.~\ref{ProposedSetup}, our proposed device setup contains two asymmetric sets of gates. The outermost two green layers represent two global gates covering the whole area of the bilayer graphene device. The inner two green layers serve as local gates, each covering half of the sample plane. The interlayer potential differences imposed by the two sets of gates are exactly opposite. The ZLMs are expected to propagate along the adjacent boundaries of the two local gates.
Compared to the conventional gating scheme, the advantage of our alternative device setup lies in the use of two global gates which is free of alignment and leaves the separation distance $L$ between two local gates  the only  parameter to adjust. Subject to the precision control during the fabrication, $L$ may fluctuate between positive and negative, corresponding respectively to the departure and overlap of the two local gates. Finely tuning the separation distance to be $L=0$, the proposed device restores the ideal setup depicted in Fig.~\ref{FIG3}(a).

\begin{figure}
\includegraphics[width=8.5cm,angle=0]{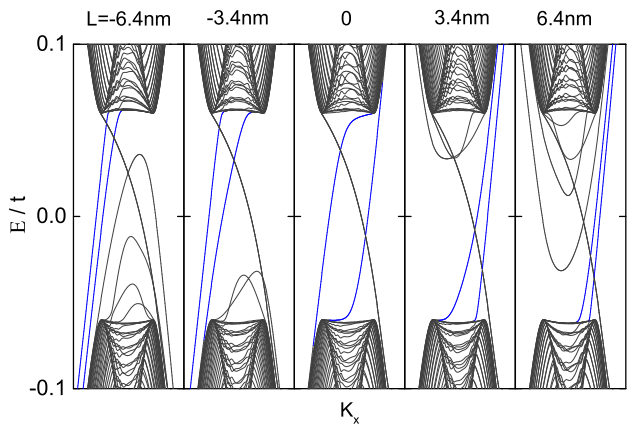}
\caption{Band structure (only shown the bands around valley K') of proposed device setup (for setup shown in Fig.~\ref{ProposedSetup}). The blue lines indicate the ZLMs. $L$ denotes the separating distance of two local gates. $L=-6.4, -3.4, 0, 3.4, 6.4$~nm in the five panels respectively. }\label{bands-proposedSetup}
\end{figure}

Fig.~\ref{bands-proposedSetup} illustrates the band evolution of our proposed device around valley K. The zero line is supposed to along the zigzag direction and the interlayer potential variations are set to be zero across the zero line.
The separation distance between the local gates is set to $L=-6.4$, $-$3.4, 0, 3.4, and 6.4~nm, respectively. For $L=0$, the ideal ZLMs appear inside the bulk band gap $\Delta$.
When $L$ increases, the bound states emerge as expected and the ZLMs are squeezed and shifted aside in momentum space. In contrast to the traditional setups where the ZLMs shrink in energy and the band dispersion is distorted,
the ZLMs produced in our device setup exist over the whole energy range of $\Delta$ and exhibit undistorted dispersion, which suggests that our alternative setup not only simplifies the experimental preparation
but also guarantees a ZLM signal of high quality.

\section{Role of Topological Defects}\label{SectionIV}
In the previous section we discussed the possible influence of the device geometry
details in the experimental realization of the ZLM in bilayer graphene.
Here we explore the effects of topological defects in the electronic structure and transport properties of the ZLM.
Engineering topological line defects and grain boundaries is a way to produce the necessary AB/BA type
stacking faults that lead to the formation of opposite valley-Hall effects and thus the zero-lines at the interface region.
AB/BA type domains have been proposed to exist theoretically in narrow graphene nanoribbons flanked by hexagonal boron nitride sheets of opposite topology \cite{nanoroad},
tilt boundaries of multilayer graphene \cite{eunah}, and they have been
experimentally observed in gated bilayer graphene at the interface region between stacking faults \cite{mceuen,NatureZLM}.
It is expected that topological line defects as well as grain boundaries can introduce the necessary
stacking faults to produce the zero-lines.
In experiments, it is known that the impurities are mainly distributed around graphene sample boundaries
and are unlikely to have destructive effects in the 1D ZLMs that are located in the interior.
Therefore, instead of focusing on weak on-site external disorder
we will consider internal topological defects (intrinsic structural defects)
that can easily appear during the growth process either in the form of point defects or grain boundaries.
Herein below, we study the effect of the topological defects on the formation of ZLMs in single layer graphene.

We use the tight-binding model Hamiltonian to investigate the effects
of the topological defects on the electronic properties of the 1D ZLMs.
Since we have shown that the ZLMs are insensitive to the boundary configurations
(i.e. zigzag, armchair, or any other type of ribbon), we will only focus on the topological
defects in the zigzag graphene ribbon as a representative case.
We assume that the hopping energies near the point defect
will approximately retain the original values of the unperturbed system.
To describe better the role of the topological defects in the 1D ZLMs we divide our discussion into two parts:
(i) periodic topological defects along the zero line;
(ii) single topological defect in a finite-sized scattering region.
\begin{figure}
\includegraphics[width=6cm,angle=0]{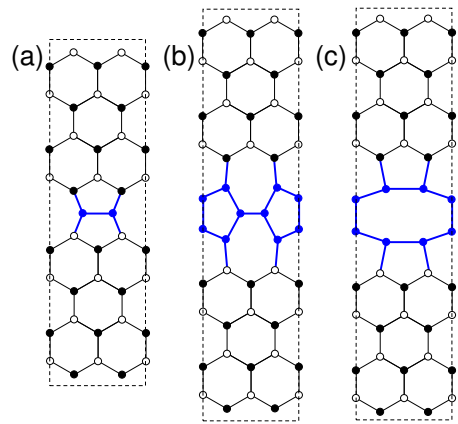}
\caption{(Color online)
Schematic plot of the unit cells of zigzag-terminated graphene ribbon in the
presence of three different kinds of topological defects.
(a) The first kind of topological defect consisting of a
line-defect with two pentagons and one octagon in the domain wall region of one unit-cell;
(b) The second kind of topological defect consisting of
two heptagons and two pentagons in the domain wall region of one unit-cell;
(c) The third kind topological defect consisting of
two pentagons and one octagon in the domain wall region of one unit-cell.
Here, the staggered AB sublattice potentials are not
considered in the domain wall region indicated with a blue color. }
\label{setup-TopologicalDefects}
\end{figure}

\subsection{Periodic Topological Defects}\label{SectionIVA}
Realistic materials can host a large variety of intrinsic topological defects.
Here in our discussion, we consider three representative types of defects
that form along the zero line highlighted in blue in Fig.~\ref{setup-TopologicalDefects}:
(a) two pentagons and one octagon; (b) two pentagons and two heptagons;
(c) two pentagons and one octagon in a form distinct form to that in Fig.~\ref{setup-TopologicalDefects}(a).
In these three kinds of topological defects, only the first one has been
experimentally observed~\cite{TologicalDefect1}, while the other two are theoretical proposals.
In our numerical calculation, we do not impose the staggered AB sublattice
potentials in the topological defects area,
because it is not possible to label the AB sublattices in the
pentagon and heptagon lattices.
\begin{figure}
\includegraphics[width=8.5cm,angle=0]{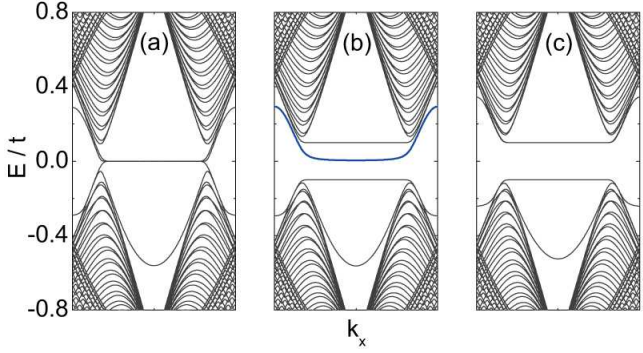}
\caption{(Color online) Band structure of zigzag-terminated graphene ribbon in the presence of the first kind of topological defects (two heptagons and one octagon in a unit-cell). (a) No AB sublattice potentials are considered; (b) AB sublattice potentials $U_0/t$=0.1 are uniformly distributed in the whole regime except the domain wall region; (c) The crystal topologies are different between the two regimes separated by the topological defects rings. The blue bands indicate the kink-state bands.} \label{ToplogicalDefect_bands1}
\end{figure}

The first kind of topological defect shown in Fig.~\ref{setup-TopologicalDefects}(a) is in a line-form.
Therefore, all the remaining atoms are still distinguishable
as A or B sublattices. For example, if we set the crystal topologies of staggered AB
sublattice potentials to be the same at both sides of the zero line,
they behave as opposite crystal topologies due to the presence of the line-defect.
This is confirmed by the resulting band structure as shown in
Fig.~\ref{ToplogicalDefect_bands1}(b), where the ZLMs are only present at one half
of the bulk band gap [\textit{i.e.}, $E/t$$\in$(0, $U_0$)].
Another interesting observation is that there exists an almost three-fold degenerate
flat bands in the absence of site potentials [see Fig.~\ref{ToplogicalDefect_bands1}(a)],
with two flat-bands appearing one for the conduction and another for
the valence band and the ZLM inside the band gap.
When we set the staggered potentials to be opposite at both sides, the line-defect also reverses
the crystal topologies and no ZLM are formed [see in Fig.~\ref{ToplogicalDefect_bands1}(c)].
To visually describe the spatial distribution of the ZLM in the presence of the first kind of topological defects,
we plot the LDOS of the ZLM injecting from the left terminal in Fig.~\ref{Defects_LDOS}(a),
which clearly shows that the ZLM is mainly located near the line-defect region.
\begin{figure}
  \includegraphics[width=8.5cm,angle=0]{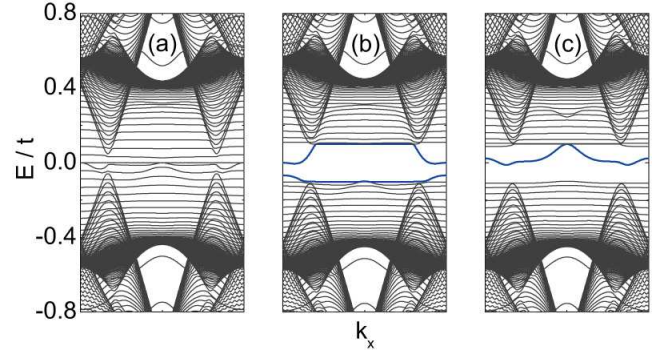}
  \caption{(Color online) Band structure of zigzag-terminated graphene ribbon in the presence of second kind topological defects (two heptagons and two pentagons in a unit-cell). (a) No AB sublattice potentials are considered; (b) AB sublattice potentials $U_0/t$=0.1 are uniformly distributed in the whole regime except the domain wall region; (c) The crystal topologies are different between the two regimes separated by the topological defects rings. The blue bands indicate the kink-state bands.} \label{ToplogicalDefect_bands2}
\end{figure}

For the second kind of topological defect, there are two heptagons and two pentagons in one unit-cell. In the absence of staggered potentials, multiple flat bands form in a wide energy range  in addition to the normal zigzag graphene bands as shown in Fig.~\ref{ToplogicalDefect_bands2}(a). When the staggered AB sublattice potentials are included, a bulk band gap opens due to the inversion symmetry breaking. From Fig.~\ref{ToplogicalDefect_bands2}(b) and Fig.~\ref{ToplogicalDefect_bands2}(c), one can observe that the ZLMs arise at both the same and opposite crystal topologies. The corresponding ZLM for the same topology is visually represented in Fig.~\ref{Defects_LDOS}(b), which also shows that the ZLM still survives in the presence of the second kind of topological defect.

\begin{figure}
\includegraphics[width=8.5cm,angle=0]{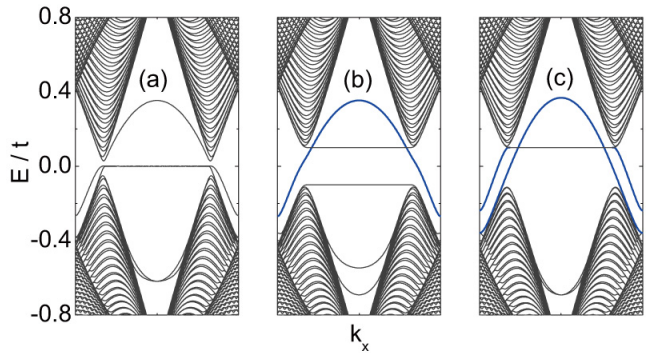}
\caption{(Color online) Band structure of zigzag-terminated graphene ribbon in the presence of third kind topological defects (two pentagons and one octagon in a unit-cell). (a) No AB sublattice potentials are considered; (b) AB sublattice potentials $U_0/t$=0.1 are uniformly distributed in the whole regime except the domain wall region; (c) The crystal topologies are different between the two regimes separated by the topological defects rings. Blue bands indicate the kink-state bands.} \label{ToplogicalDefect_bands3}
\end{figure}

\begin{figure}
\includegraphics[width=8.5cm,angle=0]{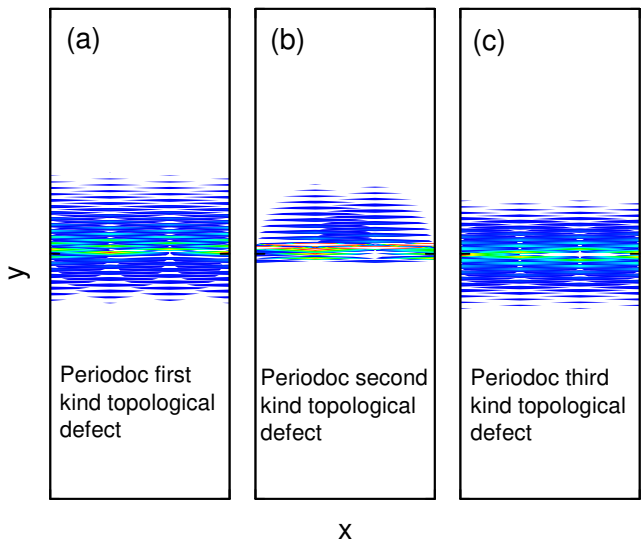}
\caption{(Color online) LDOS distribution of the kink states from the left lead for three different setups shown in Fig.~\ref{setup-TopologicalDefects}. (a) Kink state in Fig.~\ref{ToplogicalDefect_bands1}(b) from the first kind topological defect; (b) Kink state in Fig.~\ref{ToplogicalDefect_bands2}(b) from the second kind topological defect; (c) Kink state in Fig.~\ref{ToplogicalDefect_bands3}(c) from the third kind topological defect. Fermi energy is set to be $E/t=0.05$.} \label{Defects_LDOS}
\end{figure}

The unit-cells of the third and first kinds of topological defects have similar structures,
i.e. two pentagons and one octagon. However, the resulting band structures are completely different because of their difference in pattern.
In the absence of staggered potentials, the bands are similar to the normal zigzag graphene ribbon, except that there is a trivial band gap arising from the finite size effect [see Fig.~\ref{ToplogicalDefect_bands3}(a)]. When the staggered potentials are included in the calculation, we can observe gapless ZLMs both for same and opposite topology configurations.
The major difference is that there is one pair of ZLMs for the same topology [see Fig.~\ref{ToplogicalDefect_bands3}(b)], while there are two pairs of ZLMs when the topologies are opposite, as we show in Fig.~\ref{ToplogicalDefect_bands3}(c). In other words, the third kind of topological defect allows the generation of two pairs of kink states in monolayer graphene, which is not possible in the absence of topological defects. In Fig.~\ref{Defects_LDOS}(c), the LDOS distribution shows that the ZLM can still appear in the presence of the third kind of topological defect.

Because of the periodicity of the topological defects, the transport of the ZLMs is completely determined by the band structures and the conductances between left and right terminals are always exactly quantized for any Fermi energy that lies inside the bulk band gap. Based on this analysis, we conclude that the 1D ZLMs can survive even in the presence of several different kinds of periodic topological defects. However, in reality, the topological defects are not normally periodic, but appear in a random way. To complete our discussions in the following we will explore the effects of a single topological defect on the ZLMs.

\subsection{Single Topological Defect}\label{SectionIVB}
\begin{figure}
  \includegraphics[width=8.5cm,angle=0]{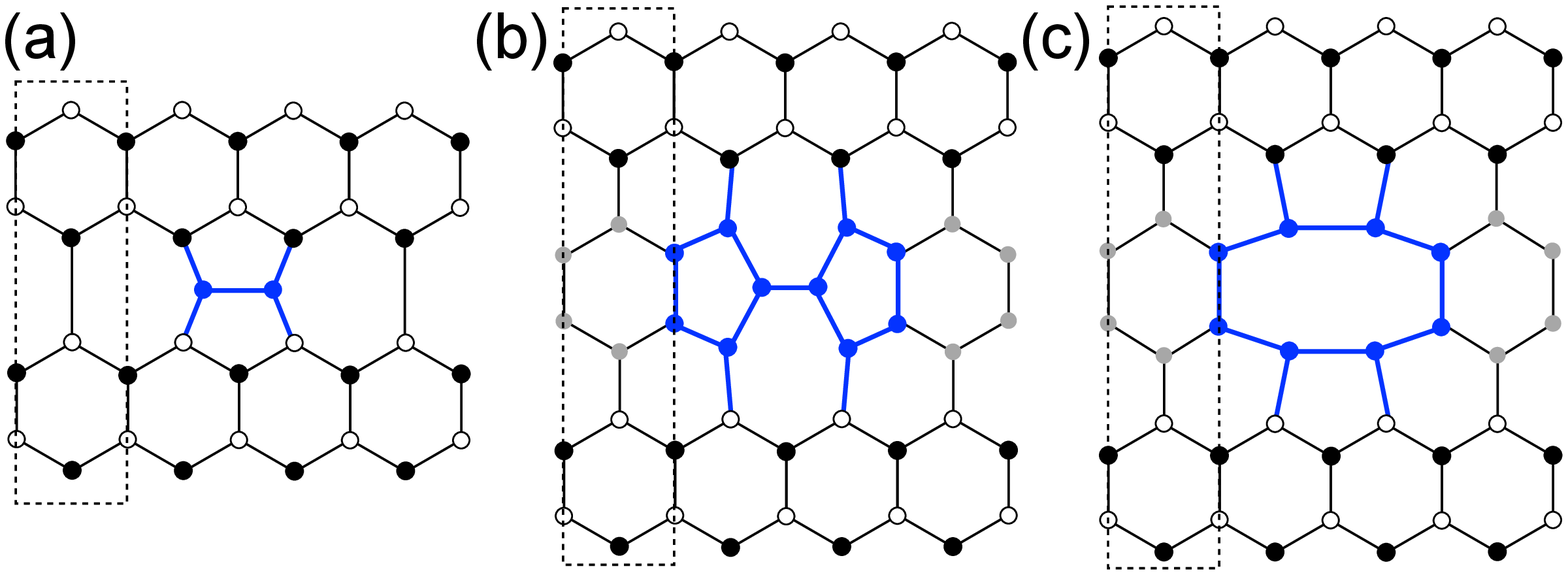}
  \caption{(Color online) Schematic plots of zigzag graphene ribbons in the presence of single topological defect. (a)-(c) stand for three different kinds of topological defects. Blue dots denote the topological defect. Grey dots denote carbon atoms in the absence of staggered AB sublattice potentials. Black and white dots are subject to staggered AB sublattice potentials. The dashed square is used to specify the periodic unit-cell along both left and right direction.} \label{setup2-TopologicalDefects}
\end{figure}

Here we would further study the effect of a single isolated topological defect on the 1D ZLMs. Similar to our analysis of periodic line defects we consider three different kinds of topological defects. In Fig.~\ref{setup2-TopologicalDefects} we show three possible configuration setups in the presence of a single isolated topological defect. The blue dots are used to represent the topological defect; grey dots represent the carbon atoms where no staggered AB sublattice potentials are applied; black and white dots have staggered AB sublattice potentials. In our numerical simulation, the studied system shows translational characteristic of the regular honeycomb lattice structures (represented with a dashed box) except the highlighted blue region. Due to the lack of the translational property of the whole lattice structures, one cannot analyze the ZLMs through plotting the corresponding band structures. Therefore, we should study the transport properties of the ZLMs in the presence of a single isolated topological defect by using the Landauer-B\"{u}ttiker formula coupled with non-equilibrium green's function technique in mesoscopic samples.

\begin{figure}
\includegraphics[width=8.5cm,angle=0]{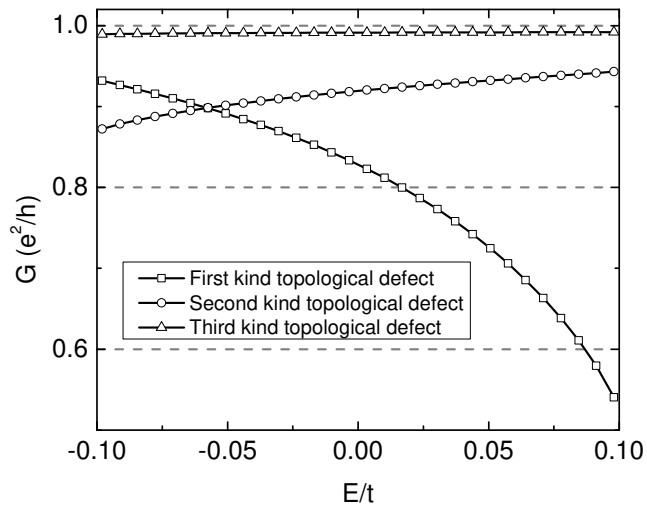}
\caption{Two-terminal conductance $G$ versus Fermi energy E for three different topological defects in a mesoscopic zigzag graphene ribbon illustrated in Fig.~\ref{setup2-TopologicalDefects}. The dimensions of the scattering region consist approximately
of $20\times320$ atoms %are $20\times320+2$, $20\times324$, and $20\times324-2$
for three different topological defects.} \label{Conductance-Defects}
\end{figure}

The two-terminal conductance $G$ is represented in Fig.~\ref{Conductance-Defects} as a function of the Fermi energy $E$ for a zigzag graphene nanoribbon in the presence of three different kinds of single isolated topological defects. %The number of atoms in the central scattering region for the three different configurations are $20\times320+2$, $20\times324$, and $20\times324-2$, respectively.
The system size in the central region correspond to an area produced by roughly $320\times20$ number of atoms in the $x$ and $y$ directions respectively.
Note that there is only one topological defect in each system. We found that: (i) the conductances are no longer symmetric about $E/t=0$; (ii) the conductance in the presence of the first kind topological defect is greatly altered; (iii) the conductances in the presence of the second kind topological defect are close to a constant of $0.9~e^2/h$; (iv) the conductances in the presence of the third kind topological defect are almost unaffected, close to the quantized value. Because the ZLMs can only emerge for the same crystal topologies for the first kind of topological defect, while it can be formed for both the same and different crystal topologies in the presence of second and third topological defects, the crystal topology can be easily and largely altered by the first kind topological defect, which results in a substantial decrease of the conductance. However, for the third kind topological defect, any crystal topology gives rise to perfect gapless kink state bands, and does not have a destructive influence in the ZLMs.

\begin{figure}
\includegraphics[width=8.5cm,angle=0]{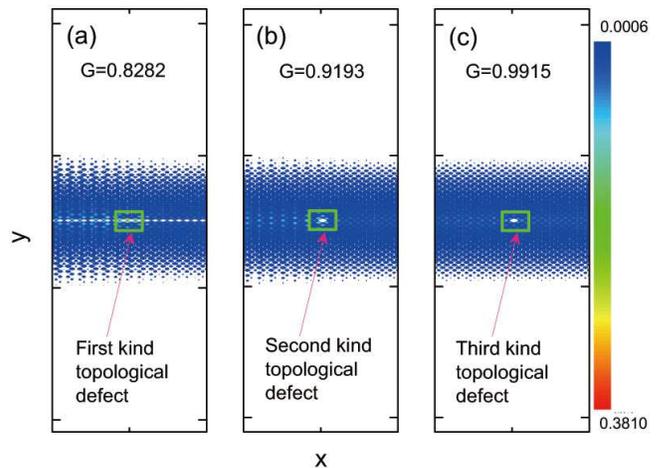}
\caption{LDOS distribution for the kink state injecting from left-lead in the presence of a single topological defect. (a) First kind topological defect; (b) Second kind topological defect; (3) Third kind topological defect. In all the three cases, Fermi energy is set to be $E/t=0$. Green squares are used to label the location of the single topological defect. Color bar represents the intensity of LDOS.} \label{SingleDefect_LDOS}
\end{figure}

From the results shown in Fig.~\ref{Conductance-Defects}, we find that these three kinds of topological defects can
have a negative influence in the preservation of the ZLM's propagation. In order to understand better how the
ZLMs are reflected in the topological defect regions we calculate
the LDOS of the ZLM coming in from the left terminal into the central scattering region in the
presence of single isolated topological defect for a zigzag graphene nanoribbon, see Fig.~\ref{SingleDefect_LDOS}.
The panels (a)-(c) correspond respectively to the three different kinds of topological defects at the fixed Fermi energy $E/t=0$.
We use the green square to indicate the location of the single topological defect. We find that although the conductances are significantly decreased for the first and second kinds of topological defects, the LDOS distributions maintain the 1D localization of a 1D ZLM in pristine samples and do not show significant deviations in shape.
The single topological defect will allow the backscattering of the right-propagating ZLM
to the counter-propagating channel located in the same zero line thanks to the strong inter-valley scattering.
We find a negligible alteration due to disorder in the spatial confinement of the ZLM produced by the kink potentials.
%in the insulating region remain the same.

To summarize this Section, we show that ZLM can still exist in systems with any kind of the three considered periodic topological defects. For single topological defect, we find that although the 1D propagating behaviour is not altered, the first kind topological defect leads to a significant degradation of the ZLM due to its high dependence on the AB sublattice symmetry. On the contrary, the ZLMs are much more robust against the remaining two topological defects. As a general statement we can conclude that the presence of the topological defects can influence the rate of backscattering of the ZLM but barely affects the confinement properties along the 1D zero-line.

\section{Conclusion}\label{SectionV}
In summary, we have studied the influence of the device geometries and topological
defects on the electronic structure and transport properties of zero-line modes in
bilayer and single layer graphene systems.
The analysis of the band evolution of the quantum valley Hall edge modes and the ZLMs in bilayer graphene ribbons
show that the edge modes of quantum valley hall effect will develop changing gaps as the ribbon orientation deviates from the zigzag direction, while the corresponding ZLMs remain all the way gapless except for the armchair direction. When the bilayer graphene ribbon evolves into the armchair type, the two pairs of ZLMs become completely mixed and leads to an abrupt opening of a small avoided gap. The ZLMs thus show more robustness than the edge modes of QVH states and the valley indices is distinguishable in non-armchair ribbons. % or zero lines.

To produce high quality zero line modes in bilayer graphene, the precise alignment of four identical gates are required in traditional device setups. We provided a theoretical analysis of two practical imperfections in the geometry of real samples: (1) The in-plane separation distance between the top/bottom gates can not be made arbitrarily small. To date the most achievable value is about 100~nm. (2) The top gates and the bottom gates can not aligned preciously. Our band structure calculations indicate that both the finite distance and the vertical misalignment of gates will suppress the ZLM signal. Specifically, the split distance will introduce localization of bulk states at the band edges and narrow the energy range of ZLMs and the misalignment between top and bottom gates will induce density imbalance of bound modes and greatly distort the dispersion curves of the ZLM. In order to observe the ZLMs  the upper limit of the split distance is about $100$~nm
for electric controlled bilayer graphene systems

To simplify the experimental preparation we propose an alternative gate setup consisting of two global gates
and two local gates. % rather than four identical gates.
In the new setup the only adjustable parameter is the in-plane alignment between the two local gates. Band structure
calculations show that high quality ZLMs  exist over a broad energy window in our proposed device setup.

We then analyzed the effects of the topological point defects in the zero-line transport.
For this purpose we simulated a single layer graphene system subject to AB staggered potentials
in the presence of point defects along the zero-line. We find that the spatial confinement of the 1D kink states
remains largely unaffected by all three types of topological defects and therefore the observability of
their local density of states through scanning microscopy probes will not be decreased.
However, the transport properties are affected by enhanced
backscattering especially when the point defects eliminate the distinction between the
AB sublattices, but the backscattering is not strong for the other types of point defects.
Our analysis of gate geometry effects and the relative robustness of the ZLM to point defects
based on lattice tight-binding and Landauer-Buttiker transport calculations
suggests optimistic prospects for the detection of ZLM signals in realistic top-bottom gated devices
with currently available gate edge alignment precision on the order of a tens of nanometers.

\acknowledgments
We thank the financial support from China Government Youth 1000-Plan Talent Program, NNSFC (11474265), and Anhui Provincial Natural Science Foundation. The supercomputing center of USTC  and the Texas Advanced Computing Center (TACC) from the University of Texas at Austin are gratefully acknowledged for the high-performance computing assistance.

\end{document}